\documentstyle{mn}
\def\spose#1{\hbox to 0pt{#1\hss}}
\def\simlt{\mathrel{\spose{\lower 3pt\hbox{$\mathchar"218$}}
     \raise 2.0pt\hbox{$\mathchar"13C$}}}
\def\simgt{\mathrel{\spose{\lower 3pt\hbox{$\mathchar"218$}}
     \raise 2.0pt\hbox{$\mathchar"13E$}}}
\newcommand{\cntr}[2]{\multicolumn{#1}{c}{#2}}
\newcommand{\alo}{\alpha_{\rm opt}}
\newcommand{\ahi}{\alpha_{\rm IR}}

\begin{document}
\title{Infrared photometry of $z \sim 1$ 3C quasars}
\author[Chris Simpson and Steve Rawlings]{Chris Simpson$^1$ and Steve
Rawlings$^2$\\
$^1$Subaru Telescope, National Astronomical Observatory of Japan, 650
N. A`oh\={o}k\={u} Place, Hilo, HI 96720, U.S.A.\\
$^2$Department of Physics, University of Oxford, Oxford OX1 3RH}
\maketitle

\begin{abstract}
We present {\it JHKL$'$\/} photometry of a complete sample of
steep-spectrum radio-loud quasars from the revised 3CR catalogue in
the redshift range $0.65 \leq z < 1.20$. After correcting for
contributions from emission lines and the host galaxies, we
investigate their spectral energy distributions (SEDs) around
$1\,\mu$m. About 75\% of the quasars are tightly grouped in the plane
of optical spectral index, $\alo$, versus near-infrared spectral
index, $\ahi$, with the median value of $\alo$ close to the canonical
value, and the median $\ahi$ slightly flatter. We conclude that the
fraction of moderately-obscured, red quasars decreases with increasing
radio power, in accordance with the `receding torus' model which can
also explain the relatively flat median near-infrared spectra of the
3CR quasars. Two of the red quasars have inverted infrared spectral
indices, and we suggest that their unusual SEDs might result from a
combination of dust-scattered and transmitted quasar light.
\end{abstract}

\section{Introduction}

It has been well-known for many years that core-dominated radio-loud
quasars appear optically brighter and bluer than lobe-dominated ones
(e.g., Jackson et al.\ 1989; Baker \& Hunstead 1995). Since powerful
radio cores are believed to result from Doppler boosting when the
radio jet is oriented close to the line of sight, this implies a
viewing angle dependence of the optical emission. Such a dependence
could be caused by Doppler boosting of optical synchrotron radiation
(Jackson et al.\ 1989), anisotropic emission from an optically-thick
accretion disc (Netzer 1985, 1987), or dust extinction and reddening
of the continuum at large viewing angles (Baker 1997). While the first
mechanism is likely to be important only in radio-loud quasars, the
others should also influence the appearance of radio-quiet quasars.
Unfortunately, the techniques for identifying radio-quiet quasars (via
optical and/or X-ray emission) are themselves affected by these
mechanisms, and cannot therefore be used to produce samples suitable
for studying them. On the other hand, samples selected at low radio
frequency are especially useful since they consist almost exclusively
of objects which are viewed at random orientations because their radio
emission is both unshadowed and unbeamed. In principle, therefore, one
can learn much about viewing angle dependencies from low-frequency
radio samples, with a view to understanding the physical mechanism(s)
responsible.

Making measurements in the region of $\lambda_{\rm rest} \approx
1\,\mu$m is also likely to prove fruitful, since the mechanisms
responsible for emission on either side of this wavelength are
believed to be distinct. At optical wavelengths (henceforth defined as
$\lambda_{\rm rest} < 1\,\mu$m), the emission is thought to arise from
the low-energy tail of an accretion disc spectrum, which could be
intrinsically anisotropic (Netzer 1987). The emission at infrared
wavelengths ($\lambda_{\rm rest} > 1\,\mu$m) is believed to be
dominated by reprocessed thermal radiation from hot dust on the inner
walls of the obscuring `torus', where any anisotropy is almost
certainly due to extrinsic effects such as dust obscuration. We can
therefore hope to discriminate between intrinsic and extrinsic sources
of anisotropy.

The 3C sample of radio sources is an obvious place to start such
investigations. Our earlier study of $z \sim 1$ 3C radio galaxies is
described in Simpson, Rawlings \& Lacy (1999, hereafter SRL), and we
present here photometry and a preliminary statistical analysis of an
identically selected sample of quasars. In Section 2, we describe our
observations and reduction method. In Section 3, we present the
results of our photometry and derive the optical and near-infrared
continuum spectral indices. In Section 4, we look for correlations
among the properties of the quasars in our sample, and investigate
unusual objects. We summarize our results in Section 5. Throughout
this paper, we adopt $H_0 = 50$\,km\,s$^{-1}$\,Mpc$^{-1}$, $q_0 =
0.5$, and $\Lambda = 0$. Our convention for spectral index, $\alpha$,
is such that the variation of flux with frequency, $S_\nu \propto
\nu^{-\alpha}$.

\section{Observations and reduction}

We have made {\it JHKL$'$\/} imaging observations of all quasars in
Laing et al.\ (1983) with $0.65 \leq z < 1.20$ which are accessible
from UKIRT (i.e., $\delta < 60^\circ$), except for 3C~454.3, which
only appears in Laing et al.\ due to its strongly Doppler-boosted
radio core. We have also imaged the $z=1.228$ quasar 3C~68.1.

With the exception of the images of 3C~68.1 and 3C~175, which were
observed on UT 1998 Sep 17, and 3C~380, which was observed on UT 1999
Mar 12, all the data in this paper were obtained during the nights of
UT 1999 Mar 6--8 with the IRCAM3 infrared array imager on UKIRT. The
{\it JHK\/} images were each constructed from a single five-point
mosaic with an exposure of 60\,s (30\,s for 3C~68.1 and 3C~175) per
position (split into multiple coadds to avoid saturation while still
achieving background-limited operation), while the $L'$ images were
produced from a number of five-point mosaics with an exposure of 24\,s
(30\,s for 3C~68.1 and 3C~175) per position (again split into multiple
coadds). The seeing, as measured from stars and the quasars themselves
at $K$-band, varied between 0.6 and 1.0 arcsec.

Each five-point mosaic was reduced separately, in the manner of SRL,
and the individual $L'$ mosaics were then averaged together to form
the final image. Many of our thermal-infrared observations were beset
by problems with the array control software, as reported by SRL. Care
was taken to ensure that only those mosaics which increased the
sensitivity of the observation were added to the final image.

Flux calibration solutions for the nights of Mar 6--8 were determined
from observations of UKIRT standard stars throughout the night. The
images of 3C~68.1, 3C~175, and 3C~380 were photometrically calibrated
using observations of standard stars taken at similar airmasses to the
quasars either immediately before or after the target observations.
Since the observations through different filters were taken within
minutes of each other (except for 3C~336, whose {\it JHK\/} and $L'$
images were taken on consecutive nights), variability is not a
concern.

\section{Photometry and spectral fitting}

\begin{figure}
\vspace*{186mm}
\includegraphics{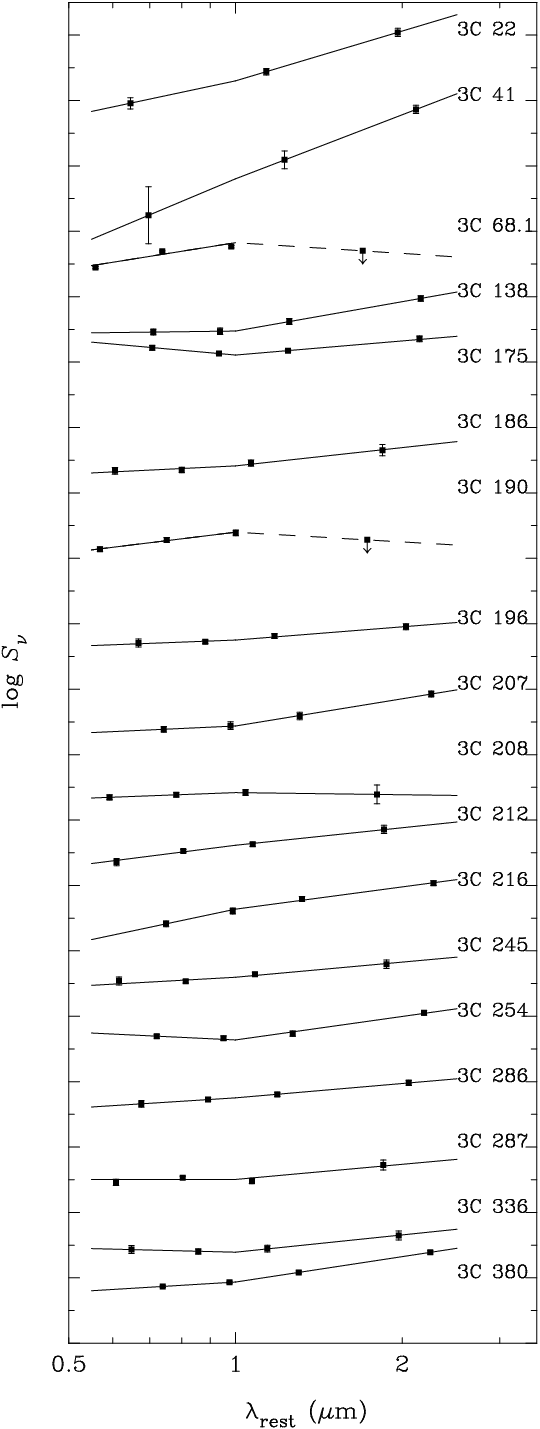}
\caption[]{Broken power law fits to the corrected quasar photometry.
Major tick marks occur every decade in flux and the labelling
indicates the major tick mark corresponding to an observed flux of
1\,mJy for each object. Dashed lines are used when only an upper limit
is available at $L'$.}
\label{fig:fits}
\end{figure}

\begin{table*}
\caption[]{Radio flux densities, Galactic extinction, and measured
photometry in a 3-arcsec aperture for the sample of quasars. All
limits are $2.5\sigma$. The Galactic $B$-band extinctions are from the
NASA/IPAC Extragalactic Database (NED). Total 5-GHz fluxes are from
Pauliny-Toth \& Kellerman (1968). References for 5-GHz core fluxes
are: (1) Bridle et al.\ (1994); (2) Cotton et al.\ (1997); (3) Hough
\& Readhead (1989); (4) L\"{u}dke et al.\ (1998); (5) Paragi, Frey \&
Sanghera (1998); (6) Pearson, Perley \& Readhead (1985).}
\label{tab:phot}
\begin{tabular}{lccr@{}lcccccr@{}l}
\hline
Name & $z$ & $S_5$ (Jy) & \cntr{2}{$S_{\rm c}$ (mJy)} & Ref & $A_B$ &
$J$ & $H$ & $K$ & \cntr{2}{$L'$} \\ \hline
3C~68.1 & 1.228 & 0.83 &    1&.1 & 1 & 0.20 &
$16.84\pm0.02$ & $15.72\pm0.01$ & $15.04\pm0.03$ &$>14.08$ &       \\
3C~138  & 0.759 & 4.16 &  460&   & 4 & 1.90 &
$16.90\pm0.03$ & $16.32\pm0.04$ & $15.38\pm0.05$ & 13.51&$\pm0.09$ \\
3C~175  & 0.768 & 0.66 &   23&.5 & 1 & 0.74 &
$14.86\pm0.02$ & $14.72\pm0.01$ & $14.10\pm0.03$ & 12.58&$\pm0.10$ \\
3C~186  & 1.063 & 0.38 &   15&   & 4 & 0.17 &
$16.84\pm0.06$ & $16.46\pm0.05$ & $15.68\pm0.06$ & 14.18&$\pm0.17$ \\
3C~190  & 1.197 & 0.82 &   73&   & 4 & 0.10 &
$17.47\pm0.04$ & $16.63\pm0.03$ & $15.84\pm0.04$ &$>14.86$ &       \\
3C~196  & 0.871 & 4.36 &    6&   & 3 & 0.16 &
$15.90\pm0.06$ & $15.61\pm0.05$ & $14.88\pm0.06$ & 13.52&$\pm0.10$ \\
3C~207  & 0.684 & 1.44 &  510&   & 3 & 0.21 &
$16.81\pm0.03$ & $16.11\pm0.04$ & $15.28\pm0.05$ & 13.56&$\pm0.09$ \\
3C~208  & 1.110 & 0.54 &   51&.0 & 1 & 0.11 &
$16.86\pm0.03$ & $16.39\pm0.03$ & $15.76\pm0.04$ & 14.70&$\pm0.25$ \\
3C~212  & 1.049 & 0.89 &  150&   & 3 & 0.08 &
$16.72\pm0.03$ & $16.07\pm0.03$ & $15.31\pm0.04$ & 13.74&$\pm0.13$ \\
3C~216  & 0.668 & 1.81 & 1050&   & 6 & 0.04 &
$16.72\pm0.03$ & $15.76\pm0.04$ & $14.86\pm0.05$ & 13.32&$\pm0.08$ \\
3C~245  & 1.029 & 1.39 &  910&   & 3 & 0.02 &
$16.25\pm0.03$ & $16.04\pm0.03$ & $15.27\pm0.04$ & 13.87&$\pm0.14$ \\
3C~254  & 0.734 & 0.79 &   19&   & 3 & 0.08 &
$16.04\pm0.03$ & $15.68\pm0.04$ & $15.01\pm0.05$ & 13.28&$\pm0.08$ \\
3C~286  & 0.849 & 7.48 &$<75$&   & 2 & 0.00 &
$15.99\pm0.03$ & $15.57\pm0.04$ & $14.88\pm0.05$ & 13.44&$\pm0.08$ \\
3C~287  & 1.055 & 3.26 &  774&   & 5 & 0.02 &
$16.50\pm0.03$ & $16.06\pm0.03$ & $15.62\pm0.04$ & 14.02&$\pm0.15$ \\
3C~336  & 0.927 & 0.69 &   20&.4 & 1 & 0.27 &
$16.51\pm0.03$ & $16.31\pm0.03$ & $15.65\pm0.04$ & 14.18&$\pm0.13$ \\
3C~380  & 0.691 & 7.50 & 2800&   & 1 & 0.20 &
$15.77\pm0.02$ & $15.11\pm0.02$ & $14.25\pm0.03$ & 12.49&$\pm0.06$ \\
\hline
\end{tabular}
\end{table*}

Photometry for each of the quasars was measured in a 3-arcsec
aperture. For the {\it JHK\/} images, the sky level was determined
from an annulus around the quasar in the usual manner, and any errors
introduced by an incorrect determination were confirmed to be
negligible. At $L'$, imperfect matching of the sky level between
frames caused by residual structure from problems with the control
software required a more complex technique to avoid large systematic
errors. A polynomial of first order in column and row number was fit
to the central region of the mosaic, excluding a 3-arcsec aperture
around the quasar, and this was used as a sky frame for calculating
the photometry. It was confirmed that changing the surface fitting
parameters varied the measured photometry by less than the photometric
uncertainty. The results are listed in Table~\ref{tab:phot}.

We correct the photometry for Galactic extinction using the extinction
law of Pei (1992), and then for contamination from the host galaxy
assuming that the hosts lie on the $K$--$z$ relation for 3C radio
galaxies determined by SRL. We assume the colours of a 4-Gyr old
stellar population from the GISSEL96 models of Bruzual \& Charlot
(1993, 1999), and find that the hosts typically contribute 10--15\% of
the measured flux. We also remove the contribution to the photometry
from the H$\alpha$ emission line assuming an equivalent width of $448
\pm 142$\,\AA\ for H$\alpha$, determined from the $0.1 \leq R < 1$
quasars of Kapahi et al.\ (1998) and Baker et al.\ (1999). This range
of radio core dominance is appropriate for our steep-spectrum
sample.

\begin{table}
\caption[]{Results of fitting a broken power law to the data of
Table~\ref{tab:phot}, after correcting for contributions from the host
galaxy and emission lines. The observed flux at $(1+z)\,\mu$m (i.e.,
$\lambda_{\rm rest} = 1\,\mu$m) and the optical and infrared spectral
indices are listed. The {\it JKL$'$\/} photometry of the reddened
quasars 3C~22 and 3C~41 from SRL has been analysed in the same manner
as the quasar photometry, and the results are included here.}
\label{tab:fits}
\begin{tabular}{lrrr@{}l}
\hline
Name & \cntr{1}{$\log S_{\rm1+z\,\mu m}$ (mJy)} & \cntr{1}{$\alo$} &
\cntr{2}{$\ahi$} \\ \hline
3C~22   & $-0.700\pm0.036$ &  $1.80\pm0.45$ &    2.53 &$\pm0.21$ \\
3C~41   & $-1.196\pm0.062$ &  $3.56\pm2.78$ &    3.25 &$\pm0.20$ \\
3C~68.1 & $-0.173\pm0.009$ &  $1.35\pm0.06$ & $<-0.55$ & \\
3C~138  & $-0.524\pm0.023$ &  $0.11\pm0.30$ &    1.49 &$\pm0.13$ \\
3C~175  &  $0.108\pm0.009$ & $-0.76\pm0.20$ &    0.72 &$\pm0.11$ \\
3C~186  & $-0.587\pm0.026$ &  $0.41\pm0.21$ &    0.93 &$\pm0.32$ \\
3C~190  & $-0.600\pm0.022$ &  $1.04\pm0.13$ & $<-0.35$ & \\
3C~196  & $-0.248\pm0.020$ &  $0.33\pm0.29$ &    0.67 &$\pm0.15$ \\
3C~207  & $-0.563\pm0.025$ &  $0.39\pm0.35$ &    1.38 &$\pm0.13$ \\ 
3C~208  & $-0.581\pm0.021$ &  $0.32\pm0.15$ &  $-0.11$&$\pm0.56$ \\
3C~212  & $-0.387\pm0.019$ &  $1.07\pm0.19$ &    0.89 &$\pm0.23$ \\
3C~216  & $-0.364\pm0.021$ &  $1.79\pm0.25$ &    1.13 &$\pm0.10$ \\
3C~245  & $-0.405\pm0.019$ &  $0.47\pm0.21$ &    0.77 &$\pm0.24$ \\
3C~254  & $-0.362\pm0.017$ & $-0.40\pm0.20$ &    1.19 &$\pm0.10$ \\
3C~286  & $-0.248\pm0.018$ &  $0.54\pm0.27$ &    0.73 &$\pm0.12$ \\
3C~287  & $-0.496\pm0.020$ &  $0.01\pm0.18$ &    0.77 &$\pm0.28$ \\
3C~336  & $-0.608\pm0.026$ & $-0.21\pm0.28$ &    0.88 &$\pm0.23$ \\
3C~380  & $-0.064\pm0.010$ &  $0.51\pm0.13$ &    1.29 &$\pm0.07$ \\
\hline
\end{tabular}
\end{table}

Following the usual procedure (e.g., Neugebauer et al.\ 1987), we fit
the SED of each quasar as two power-laws which meet at $1\,\mu$m. We
denote the spectral index of the power-law in the rest-frame optical
as $\alo$, and that in the infrared as $\ahi$. To the new data
presented in this paper, we add SRL's {\it JKL$'$\/} photometry of
3C~22 and 3C~41 which, as those authors showed, are more correctly
classified as quasars than radio galaxies. It is of course important
to remember that the distinction between ``quasars'' and ``radio
galaxies'' is not clear-cut and objects such as 3C~65 and 3C~265 could
also be considered as quasars.  As the quality of available data
improves, perhaps probing dust columns larger than $A_V \approx
15$\,mag, more ``radio galaxies'' are likely to be reclassified as
quasars. For the purposes of this study, however, we are interested in
sources whose continua at $\lambda_{\rm rest} \sim 1\,\mu$m are
predominantly non-stellar. The results of our fitting are listed in
Table~\ref{tab:fits} and shown graphically in Fig.~\ref{fig:fits}.

\section{Analysis}

Since 3C~68.1 is not a member of the complete sample, we exclude it
from the following analyses, although we continue to indicate its
location on plots. We likewise exclude 3C~22 and 3C~41, since their
observed properties are highly influenced by significant internal
extinction.

\subsection{Spectral indices}

\begin{figure}
\vspace*{91mm}
\includegraphics{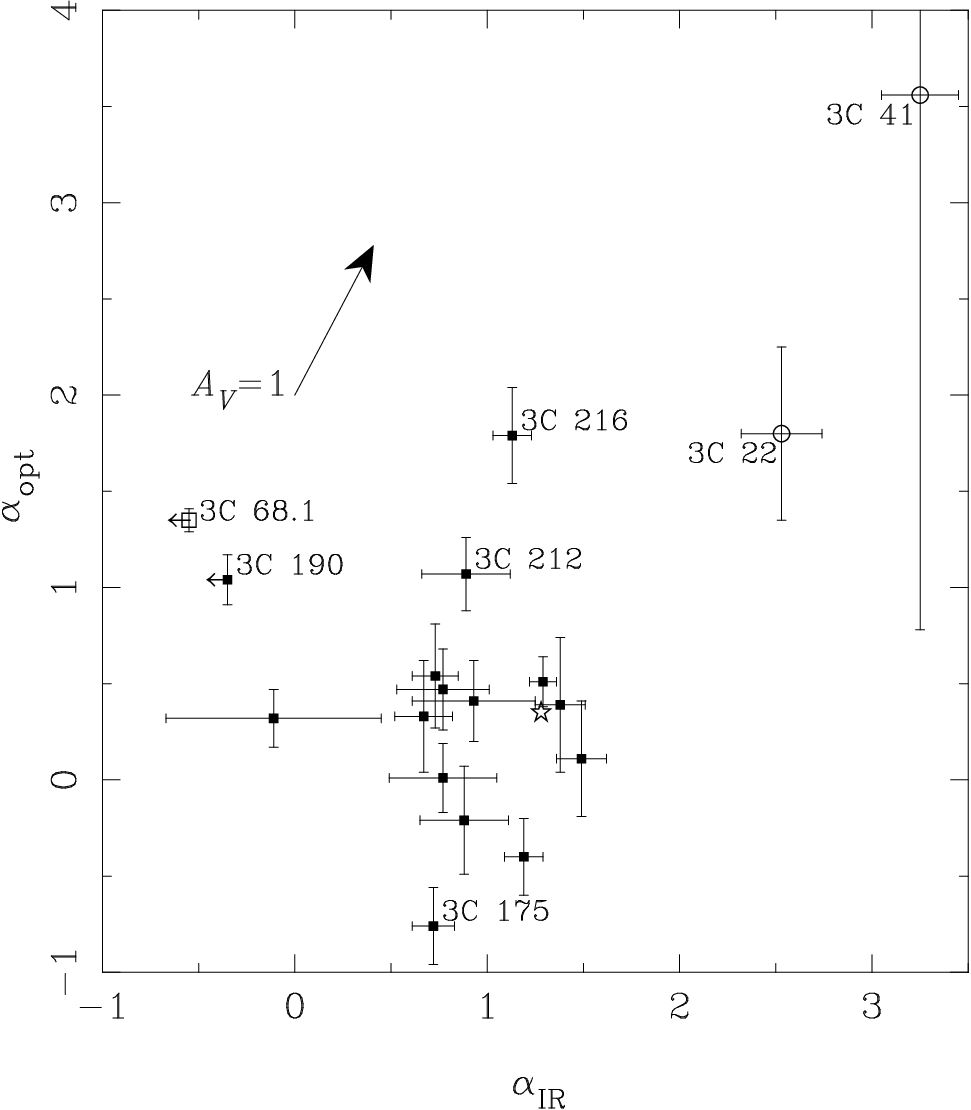}
\caption[]{Optical versus infrared spectral index for the broken power
law fits shown in Fig.~\ref{fig:fits}. Objects not in the complete
sample are shown with open symbols (an open square for the quasar
3C~68.1, open circles for the `radio galaxies' 3C~22 and 3C~41).
Objects discussed in the text are labelled. The arrow indicates the
approximate effect of one magnitude of internal reddening (the exact
translation is dependent on redshift and the relative uncertainties in
the individual flux measurements). The star shows the median spectral
indices for the quasar sample of Neugebauer et al.\ (1987).}
\label{fig:alpha}
\end{figure}

In Fig.~\ref{fig:alpha} we show the distribution of spectral indices.
Analysis reveals no significant correlation between $\alo$ and $\ahi$
(Table~\ref{tab:correl}). Of particular note are the six objects with
$\alo > 1$. These include the two `radio galaxies' 3C~22 and 3C~41,
and all four of the red objects studied by Smith \& Spinrad (1980)
common to Laing et al.\ (1983). Of these objects, two (3C~212 and
3C~216) appear to be lightly-reddened ($A_V \approx 1$, less than
3C~22 and 3C~41) quasars, while two (3C~68.1 and 3C~190) have strongly
inverted IR spectra; we shall discuss these in Section 4.3.

\begin{table}
\caption[]{Correlation analysis for the complete sample. For each pair
of variables the top line gives the value of Spearman's rank
correlation coefficient, while the bottom line gives the probability
of this being due to chance if the variables were uncorrelated. The
radio luminosity at 408\,MHz has been computed from the 178-MHz flux
densities and radio spectral indices listed in Laing et al.\ (1983).
The core-dominance parameter $R$ has been calculated at 10\,GHz using
the observed 5-GHz properties listed in Table~\ref{tab:phot} and
assuming a high-frequency spectral index $\alpha = 0.75$. These
frequencies have been chosen to minimize the $k$-corrections for our
$z\sim1$ sample.}
\label{tab:correl}
\begin{center}
\begin{tabular}{lrrrr}
\hline
 & \cntr{1}{$\ahi$} & \cntr{1}{$L_{1\,\mu{\rm m}}$} &
\cntr{1}{$L_{\rm408\,MHz}$} & $R_{\rm10\,GHz}$ \\ \hline
$\alo$ & 0.011 & 0.075 & 0.261 & 0.382 \\
& 0.970 & 0.791 & 0.348 & 0.264 \\
$\ahi$ & & $-0.504$ & $-0.504$ & 0.375 \\
& & 0.057 & 0.057 & 0.265 \\
$L_{1\,\mu{\rm m}}$ & & & 0.575 & $-0.139$ \\
& & & 0.025 & 0.621 \\
\hline
\end{tabular}
\end{center}
\end{table}

We remark first on the tight grouping of about 75\% of the complete
quasar sample in Fig.~\ref{fig:alpha}. This grouping is reminiscent of
the situation for bright optically-selected quasars (e.g.\ Neugebauer
et al.\ 1987), and suggests that, as has been shown for such objects,
reddening has both a low mean and dispersion (e.g., Rowan-Robinson
1995). For optically-selected samples this could be a selection
effect, but in the case of the radio-selected 3C quasars it proves
that most of the quasars are unobscured along the line of sight. When
we also consider the results of SRL, we conclude that the transition
region over which powerful steep-spectrum radio sources have
moderate extinction covers a small solid angle. These moderately
extinguished objects (the `red quasars') are therefore much rarer than
at lower radio powers (e.g., Baker 1997).  This is, of course,
consistent with the predictions of the `receding torus' model (e.g.,
Lawrence 1991).

The median optical and infrared spectral indices in our complete
sample are 0.39 and 0.88, respectively. In the optically-selected,
predominantly low-redshift (79/104 sources have $z<0.5$) sample of
Neugebauer et al.\ (1987), the medians are 0.35 and 1.28,
respectively. While the optical spectral indices are similar, our
sample of quasars have significantly flatter infrared spectra; the
probability of 12 of our sources having flatter infrared spectra than
the median from Neugebauer et al., if the distributions were similar,
is $<2$\%.\footnote{The median $K-L'$ colour, before any corrections are
applied, is 1.54, corresponding to $\alpha=0.84$, so this result is
insensitive to the corrections outlined in Section 3.} 

Since the near-infrared emission in quasars is believed to be thermal
emission from hot dust, flatter near-IR spectra can be explained by a
decrease in the dust emission, relative to the optical luminosity, in
more luminous quasars. If the fraction of radiation incident on the
inner wall of the `torus' reprocessed by dust is constant between
objects, the strength of the near-IR excess relative to the underlying
power law is simply proportional to the solid angle subtended by the
`torus'. In terms of the receding torus model (Lawrence 1991; see also
Fig.~1 of Simpson 1998), this angle is $\sim \pi h/r$, where $r$ and
$h$ are the inner radius and half-height of the torus, respectively.
Since $h$ is assumed to be constant, $r \propto L^{0.5}$, and the
median optical luminosity of our sample is 2.7 times that of
Neugebauer et al.'s, we would expect the near-IR excess to be $\sim
40$\% weaker in our sample. This is in fair agreement with the 50\%
reduction determined from the difference in the median spectral
indices. This scenario is further supported by the fact that $\ahi$ is
anti-correlated with both 1-$\mu$m and radio luminosity, while $\alo$
is not (Table~\ref{tab:correl}). The fact that $\alo$ and
$L_{1\,\mu{\rm m}}$ are uncorrelated also demonstrates that dust is
unimportant in most of our sources, since its dual
reddening/extinction effects would produce an anti-correlation.

\subsection{Comparison with radio properties}

\begin{figure}
\vspace*{108mm}
\includegraphics{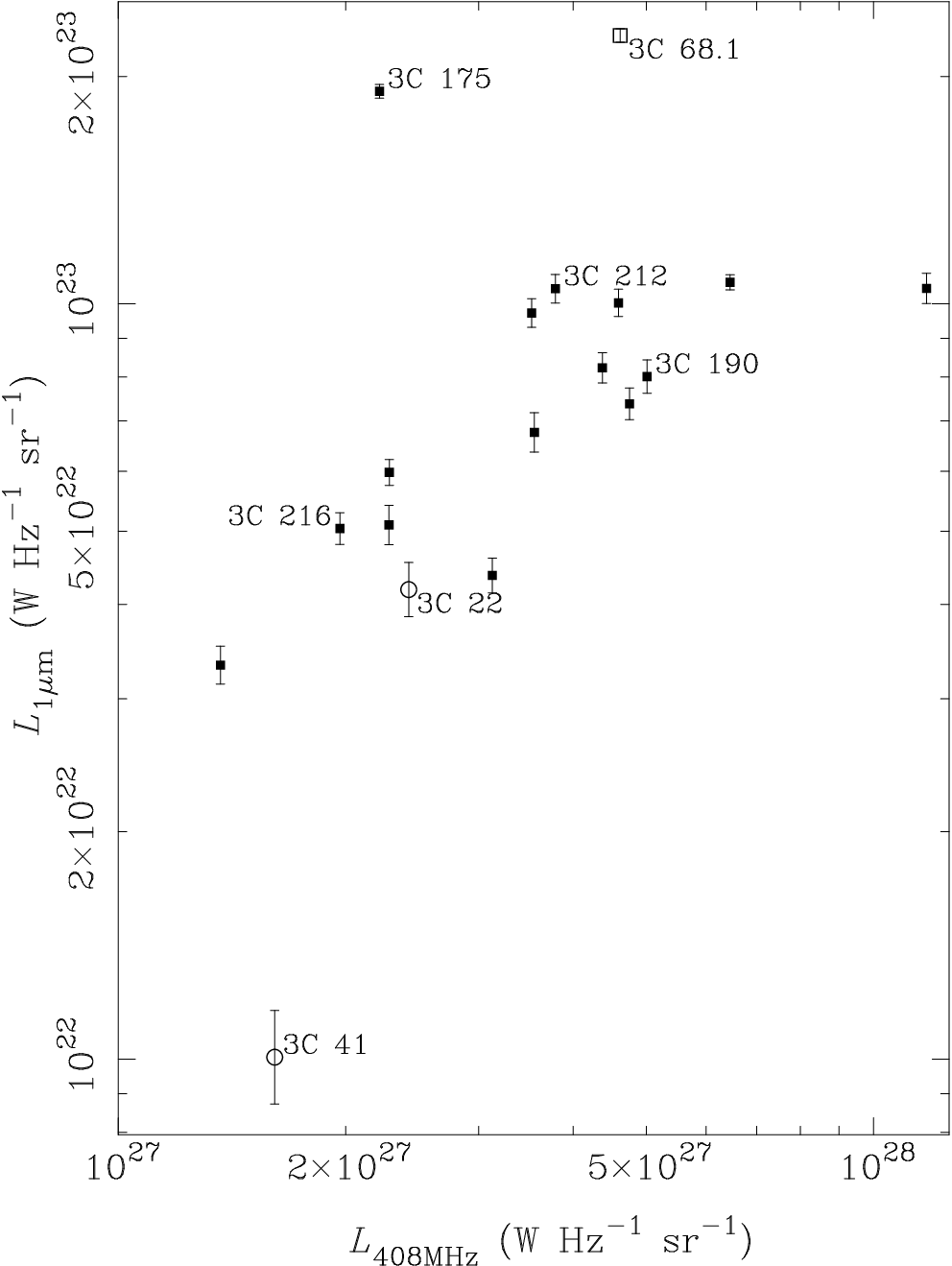}
\caption[]{Rest-frame continuum luminosities at 1\,$\mu$m and
408\,MHz. A correlation is apparent at $>97$\% significance. Symbols
are the same as in Fig.~\ref{fig:alpha}.}
\label{fig:radir}
\end{figure}

Fig.~\ref{fig:radir} shows the correlation between 1-$\mu$m luminosity
and 408-MHz radio luminosity. Since 1\,$\mu$m is a local minimum in
the typical quasar SED between the `big blue bump' from the accretion
disc at shorter wavelengths and the `big red bump' from thermal dust
emission at longer wavelengths, the luminosity here will be less
affected by the extreme properties of one or other of these bumps. The
correlation between these two properties is significant at $>97$\%
confidence (Table~\ref{tab:correl}). A least-squares fit produces a
slope of $0.41\pm0.20$, very similar to the value of $0.4\pm0.1$ found
by Willott et al.\ (1998) for the correlation between $B$-band and
radio luminosity, and slightly flatter than the $0.6\pm0.1$ result of
the original paper on this correlation by Serjeant et al.\ (1998). We
would expect a similar slope for the 1\,$\mu$m--radio correlation
since the small scatter in $\alo$ shown in Fig.~\ref{fig:alpha}
ensures that the luminosities at 4400\,\AA\ and 1\,$\mu$m are very
tightly correlated. Serjeant et al.\ claim that the tight
optical--radio correlation provides evidence for a direct link between
accretion and the fuelling of the radio jets, while Willott et al.\
suggest a possible cause for the discrepant slopes.

Finally, we compare our derived optical/near-infrared spectral indices
with the degree of radio core dominance. We compute the value of the
core-dominance parameter, $R$ (e.g., Hine \& Scheuer 1980), at a
rest-frame frequency of 10\,GHz, rather than the more common 5\,GHz,
since 5\,GHz is the most common {\em observing\/} frequency at which
core measurements are made, and we therefore need to make only small
$k$-corrections for our objects at $z\sim1$. As Table~\ref{tab:correl}
shows, neither spectral index is correlated with $R$. This tells us
that the mechanism responsible for causing the more highly-inclined
quasars to appear redder in the optical (e.g., Baker \& Hunstead 1995;
Baker 1997) is not operating for the most luminous (3C) quasars. We
mention two possible causes of this result. First, our photometry does
not probe wavelengths below $\lambda_{\rm rest} \simlt 0.5\,\mu$m, so
we are not as sensitive to reddening effects (or the increased
prominence of a UV-bright component, such as from an accretion disc)
as observations which probe shorter wavelengths. Second, and we
believe more importantly, the 3C quasars have extreme radio
luminosity, and hence (e.g., Fig.~\ref{fig:radir}; Serjeant et al.\
1998) extreme optical luminosity. According to the receding torus
model discussed in Section 4.1, the fraction of lightly-reddened
quasars should be much lower in this sample than in samples containing
objects with significantly lower radio luminosity. Comparing our
results on $z \sim 1$ 3C objects (SRL, this paper) with lower-redshift
3C sources (Hill, Goodrich \& DePoy 1996), and with coeval but less
luminous quasars (Baker 1997), it seems clear that the fraction of
dusty red quasars declines significantly with increasing radio
luminosity.

\subsection{Inverted-infrared spectra: 3C~68.1 and 3C~190}
\label{sec:invert}

We return to the two objects with apparently inverted near-infrared
spectra, namely 3C~68.1 and 3C~190. These inverted spectra cannot
result from erroneous corrections to the photometry. In the case of
3C~68.1, even if we assume that the host galaxy has a Rayleigh-Jeans
spectrum, it would need to be $\sim 2\sigma$ brighter than the mean
$K$--$z$ relation to even produce a flat IR spectrum. Alternatively,
the contamination from emission lines to the $K$-band magnitude would
need to be $> 25$\%, which is highly implausible given that the
strongest line in the highly-transmissive part of the filter is the
narrow line of [S~III]~$\lambda$9532. We note that Rieke, Lebofsky \&
Wi\'{s}nikewski (1982) detected 3C~68.1 at $L$ (3.4\,$\mu$m) with a
flux of $0.72\pm0.17$\,mJy, which compares to our upper limit of
0.60\,mJy at this wavelength (assuming a power law between $K$ and
$L'$). The two results are therefore not in conflict if 3C~68.1 has a
true flux close to our upper limit, but note that 3C~68.1 appears to
be highly variable in the infrared over a timescale of a few years
(Rieke et al.\ 1982; Stein \& Sitko 1984).

If the spectra of 3C~68.1 and 3C~190 are merely intrinsically blue
from UV to IR wavelengths but heavily-reddened, we would expect to
find other objects with similar intrinsic spectra, but unreddened.
Therefore, we suspect there is a link between the red optical and blue
infrared spectra of these two objects. Scattering by small dust grains
seems plausible since it can produce a spectrum which is bluer than
that of the incident radiation, but the dust will substantially redden
the spectrum in the rest-frame UV. This effect is demonstrated by the
modelling and spectral fits of Manzini \& di Serego Alighieri (1996);
although these authors do not explicitly consider the region longward
of 1\,$\mu$m, the scattering efficiency of dust grains is expected to
decrease monotonically with wavelength. In support of this idea,
3C~68.1 is known to be highly polarized at optical wavelengths (Moore
\& Stockman 1981; Brotherton et al.\ 1998), and a combination of a
heavily-reddened ($A_V \approx 1.7$) $\alpha \approx 0$ power law
(note that the harder power law and larger reddening compared to
Brotherton et al.'s analysis is a result of our additional infrared
data) and a more lightly-reddened ($A_V \approx 1.2$), dust-scattered
version of the same spectrum (after Manzini \& di Serego Alighieri
1996) can approximately reproduce the observed optical--infrared SED
and optical polarization. This model predicts an observed $K$-band
polarization of $\sim 6$\%. Given the almost identical locations of
the two objects in Fig.~\ref{fig:alpha}, we predict that 3C~190 should
have a similar degree of optical polarization although Moore \&
Stockman's (1984) measurement of $P = 5.34 \pm 2.81$\% is
inconclusive.

Interestingly, 3C~68.1 is also an outlier in Fig.~\ref{fig:radir},
being anomalously luminous at $1\,\mu$m given its radio luminosity.
The same is true of 3C~175, which has an intrinsically hard optical
spectrum (Table~\ref{tab:fits}), much as we have inferred for 3C~68.1.
It is worth noting that a correction for an intrinsic extinction of
$A_V \approx 2$ towards 3C~68.1 would double its 1-$\mu$m luminosity,
and a similar correction applied to 3C~190 would cause it to also
become an outlier in Fig.~\ref{fig:radir}. (Note that a correction for
the intrinsic extinction towards 3C~41 would cause it to lie on the
correlation.) Since these objects are not strongly core-dominated,
their luminous, intrinsically blue spectra are very unlikely to be the
result of orientation effects such as were discussed earlier. Instead,
we speculate that these objects are currently undergoing periods of
anomalously high accretion which is temporarily boosting their optical
luminosities. The known variability of 3C~68.1 at rest-frame
wavelengths of $1\,\mu$m is in accord with this picture.

\section{Summary}

We have presented {\it JHKL$'$\/} photometry of a complete sample of
$z \sim 1$ quasars from the revised 3CR catalogue and investigated
their spectral energy distributions in the region of 1\,$\mu$m. We
find evidence to support the `receding torus' model from the
near-infrared spectral indices of quasars in our sample, which are
both bluer than those of lower-luminosity quasar samples, and are
correlated with luminosity in the sense that the most luminous objects
within our sample have bluer near-IR spectra. In addition, the lack of
a correlation between optical spectral index and quasar orientation or
luminosity is also consistent with the predictions of this model. We
find that two of the red quasars in our sample possess unusual
near-infrared SEDs, which we attribute to a substantial dust-scattered
component, and which supports the idea that the red quasar phenomenon
is a result of dust and not due to some intrinsic properties of the
quasars themselves.

\section*{Acknowledgments}

The United Kingdom Infrared Telescope is operated by the Joint
Astronomy Centre on behalf of the U. K. Particle Physics and Astronomy
Research Council. This research has made use of the NASA/IPAC
Extragalactic Database (NED) which is operated by the Jet Propulsion
Laboratory, California Institute of Technology, under contract with
the National Aeronautics and Space Administration. We are grateful to
the anonymous referee for his/her comments.

\end{document}